\documentclass[aps,prl,twocolumn,showpacs]{revtex4}

\usepackage{graphicx}
\usepackage{dcolumn,comment}

\def\p1{\partial_1}
\def\p2{\partial_2}
\def\p3{\partial_3}

\begin{document}

\title{Meta-materials with negative refractive index at Optical Frequencies}

\author{S. Anantha Ramakrishna$^{*,\dagger}$, Sangeeta Chakrabarti$^*$ and Olivier J.F. Martin$^\dagger$}

\affiliation{$^*$ Department of Physics, Indian Institute of Technology, Kanpur 208016, India \\
$^\dagger$ Laboratory for Nanophotonics and Metrology, The Swiss Federal Institute of Technology, 1015 Lausanne, Switzerland}

\begin{abstract}
We demonstrate that arrays of split ring resonators (SRR) can have negative refractive index  
at optical frequencies ($\sim$1.4 eV). Our calculations reveal that the electric fields of radiation 
interact strongly with even symmetric SRR at optical frequencies as the size of the SRR becomes of the 
order of the wavelength ($\sim \lambda /3$) for practicably realizable structures. We also demonstrate by calculations the 
focussing of a line source by a flat slab of a (2D) SRR medium. The negative refractive index here is 
related to the plasmonic excitations of the SRR and cannot be directly explained by the usual paradigm 
of negative dielectric permittivity and negative magnetic permeability in these limits when homogenization 
breaks down. 
\end{abstract} 
\maketitle

Negative refractive index materials (NRM) were first proposed by Veselago\cite{veselago} as materials having a negative dielectric permittivity ($\varepsilon <0$) and a negative magnetic permeability ($\mu <0$). NRM have grown in popularity since practical designs for structured metamaterials  with $\varepsilon <0$\cite{pendry96} and 
$\mu <0$\cite{pendry_IEEE} at any frequency became available and were implemented at microwave frequencies\cite{NRM_expt}. A large number of counterintuitive effects for wave propagation in NRM including the possibility of super-lenses\cite{pendry00} with sub-wavelength resolution have been predicted(See \cite{sar_rop} for a recent review).  

One of the main challenges today is to obtain NRM at optical frequencies. Although several noble and alkali metals have $\varepsilon <0$ at  ultra-violet and optical frequencies, magnetic properties due atomic or molecular orbital currents or electronic spin tend to be negligible at optical frequencies. It has been shown both theoretically\cite{obrien_jpc} and experimentally\cite{soukoulis_srrexpt} that the magnetic properties of Split ring resonator (SRR) media tend to tail off at high frequencies due to the plasma-like behaviour of the metal of which they are composed of. Essentially there is an extra {\it inertial} inductance for the SRR due to the finite electronic mass and simply scaling down the SRR size does not help to obtain negative magnetic permeability. Novel ideas to overcome this limitation have been put forth such as modified SRR with more splits\cite{obrien_prb}, coupled parallel wire-pairs\cite{shalaev} and plate pairs\cite{dolling_OL} and specifically arranged plasmonic nano-particles\cite{engheta_OE}.  One common problem with all these proposals is that for practicably realizable sizes of the structures, the lengthscale of the structure becomes a sizeable fraction of the wavelength of radiation. Thus homogenization and assignment of effective material parameters is often a problematic issue.

Here we show that an array of suitably modified SRRs can exhibit negative phase velocity or negative refractive index  at optical frequencies. The SRR, which are symmetric and have no bianisotropy in the quasi-static limit, are shown to intensely interact with the electric field of the radiation.  This interaction turns out to be crucial for the negative refractive index. We also demonstrate that an array of these SRR can focus a source via negative refraction or via negative dielectric permittivity depending on the orientation of the SRR. 

Consider p-polarized radiation normally incident on an array of the two dimensional SRR of silver. We consider four possible designs for the SRR  as shown in Fig. 1(a, b, c, d).                                                                     
Invariance in the $y$ direction is assumed. Due to similiar sizes, the geometric inductance and the capacitance of the structures are similiar which are expected to have similiar magnetic properties. In fact, structure (c) is identical to structure (b) but only rotated by $\pi/2$, while structure (d), the typical plate pairs of\cite{dolling_OL} is the same as structure(c) with the shorter legs absent. In Fig. 1(e, f, g, h ) we present the band structure diagrams where both the real (solid squares) and imaginary parts (hollow circles) of the wave-vector at normal incidence are shown. For comparision, we have kept the same colour (black or red) for the real and imaginary parts of the eigenvalues with positive or negative real parts, respectively.  In Fig. 1(i, j, k, l) we show the transmittance and reflectance for a slab composed of four layers of such SRRs respectively. These calculations were carried out by a refined version of the PHOTON codes based on the transfer matrix method\cite{pendry_tmm}. The experimental values for the dielectric constant of silver\cite{johnson_christy} were used in the simulations for the SRR material. The eigenvalues are chosen so that the solutions in the dissipative medium correspond to a decaying wave in the forward direction (along the Poynting vector).

\begin{figure*}[tb]
\begin{center}
\includegraphics[angle=-0,width=1.8\columnwidth]{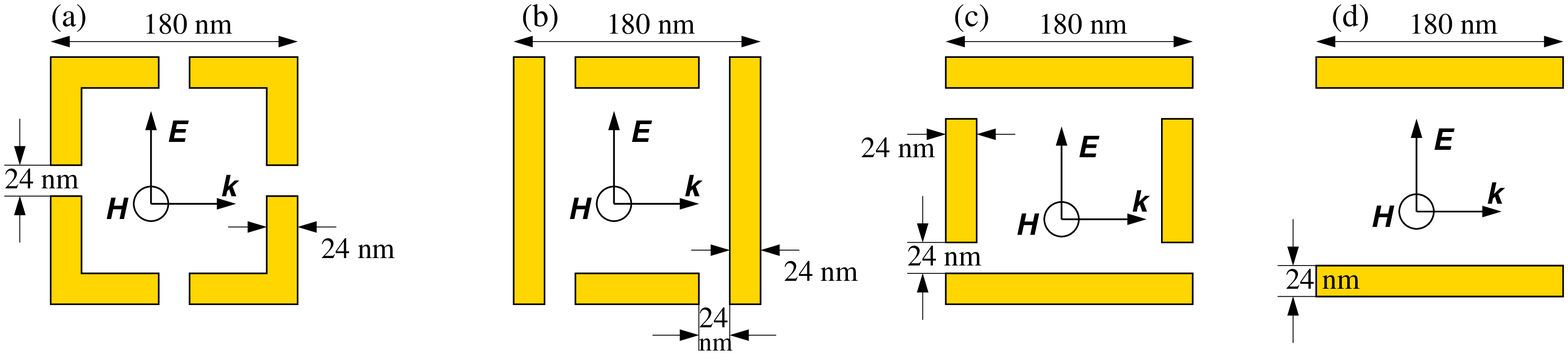}\\

\includegraphics[angle=-0,width=0.45\columnwidth]{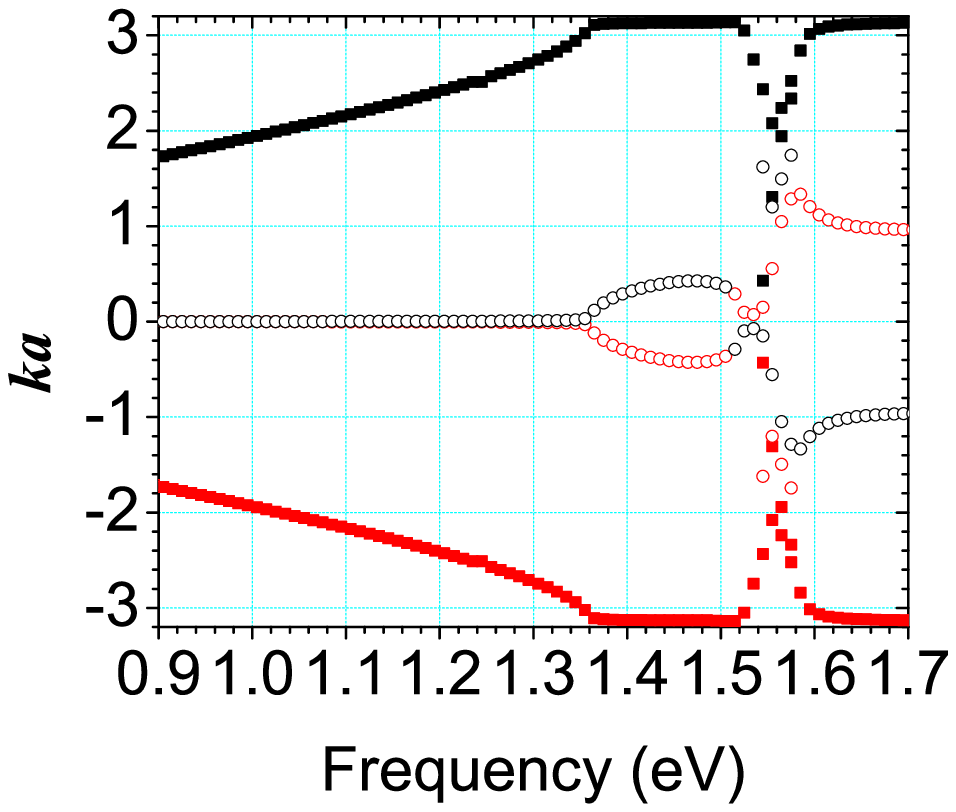}
\includegraphics[angle=-0,width=0.45\columnwidth]{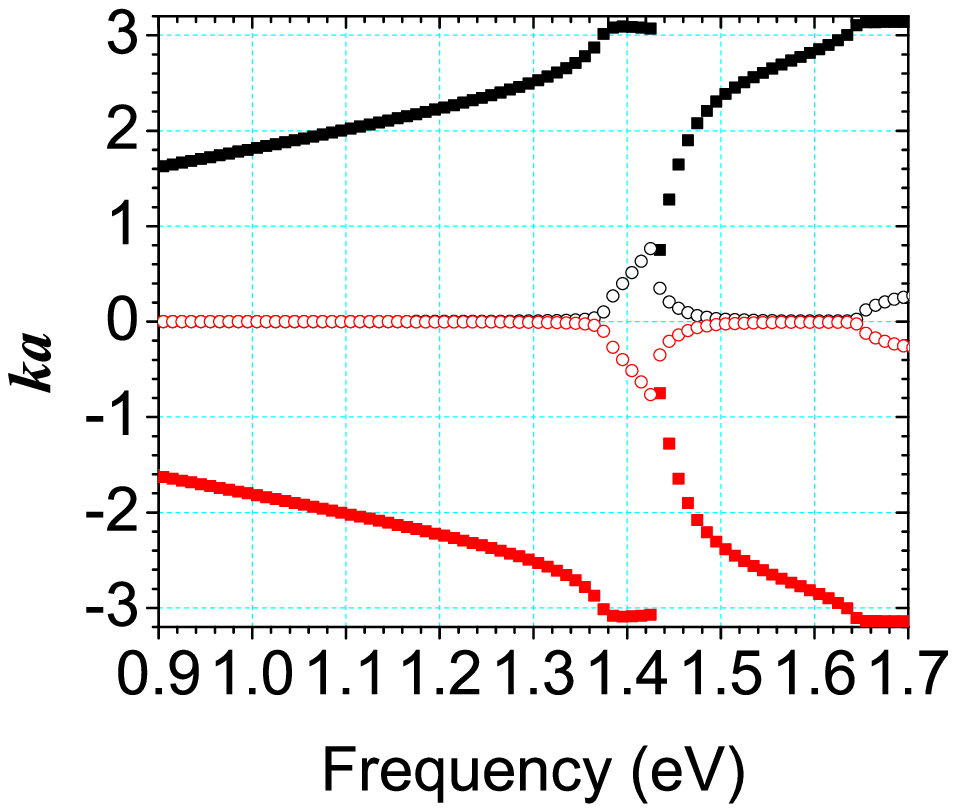}
\includegraphics[angle=-0,width=0.45\columnwidth]{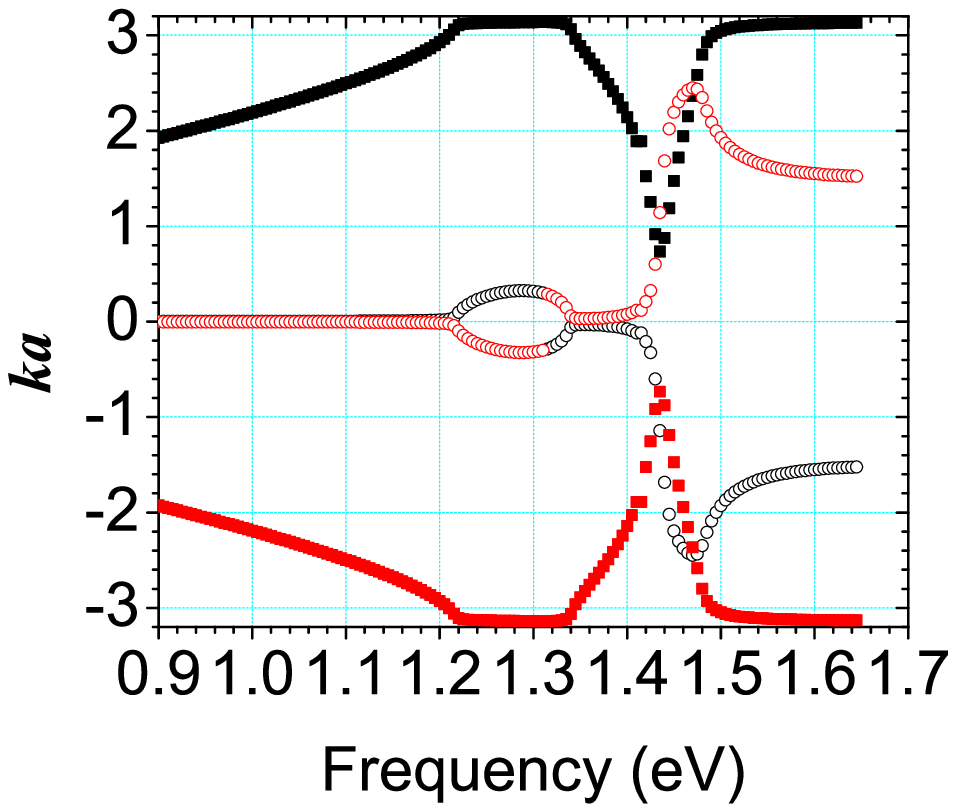}
\includegraphics[angle=-0,width=0.45\columnwidth]{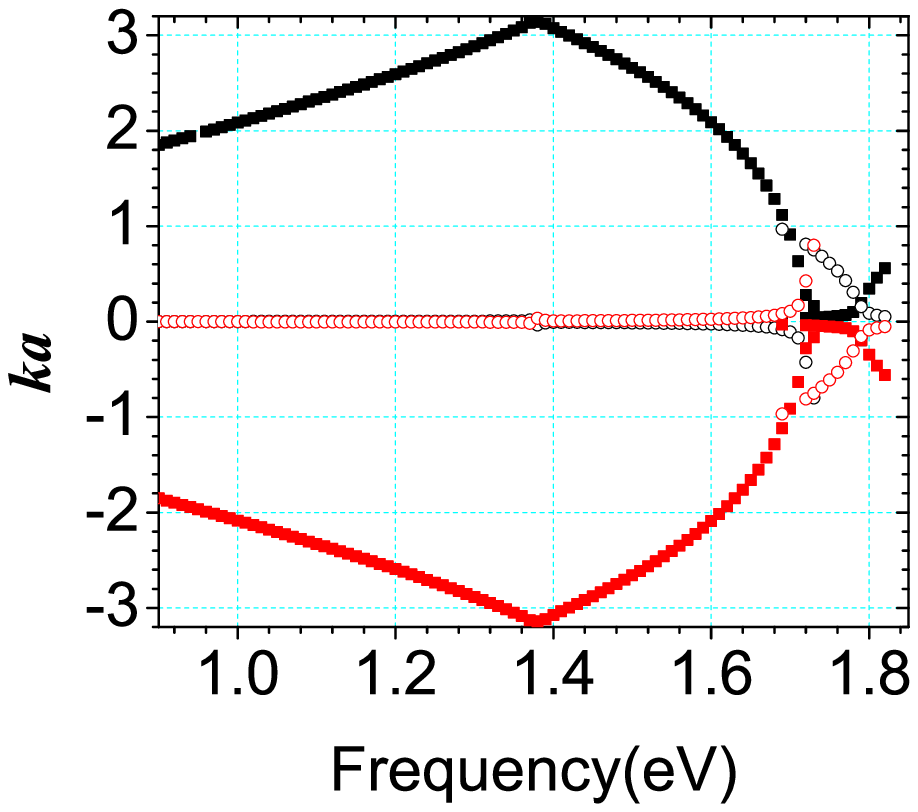}\\
\includegraphics[angle=-0,width=0.45\columnwidth]{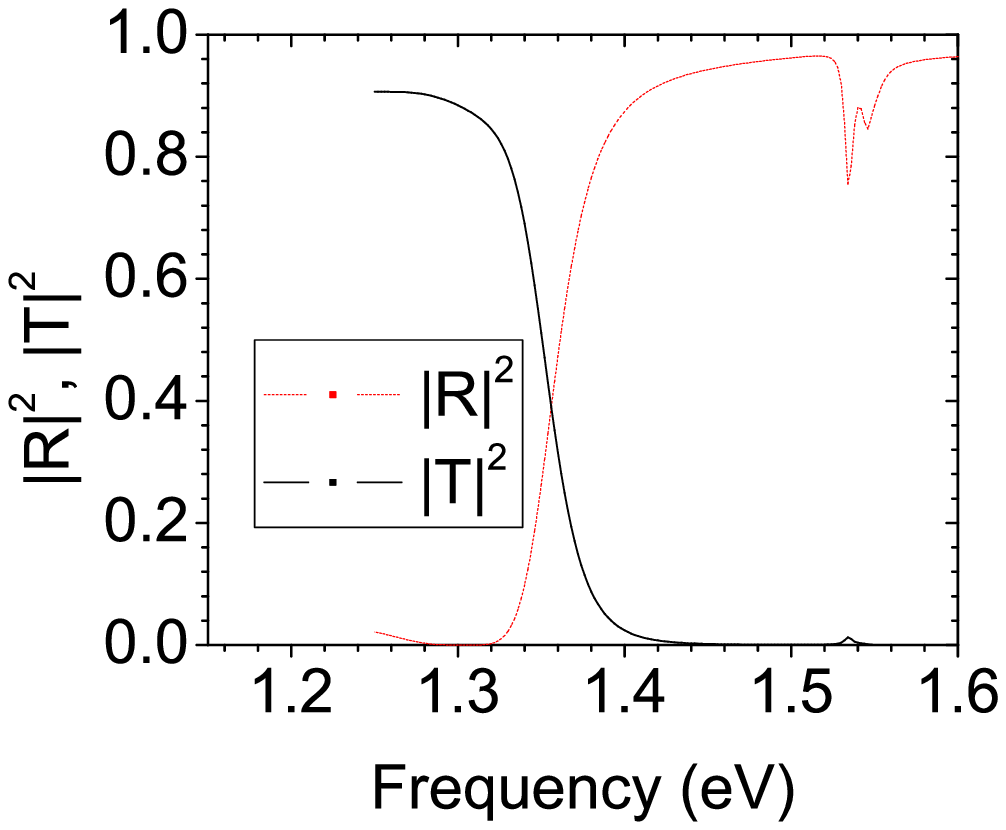}
\includegraphics[angle=-0,width=0.45\columnwidth]{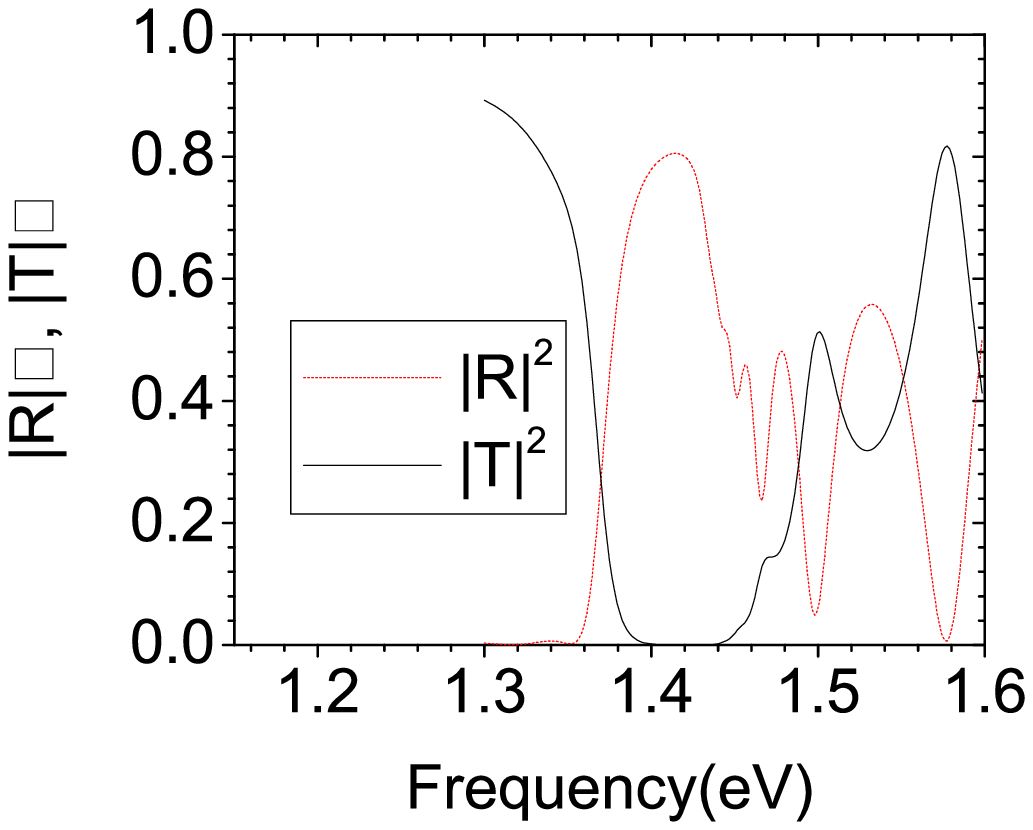}
\includegraphics[angle=-0,width=0.45\columnwidth]{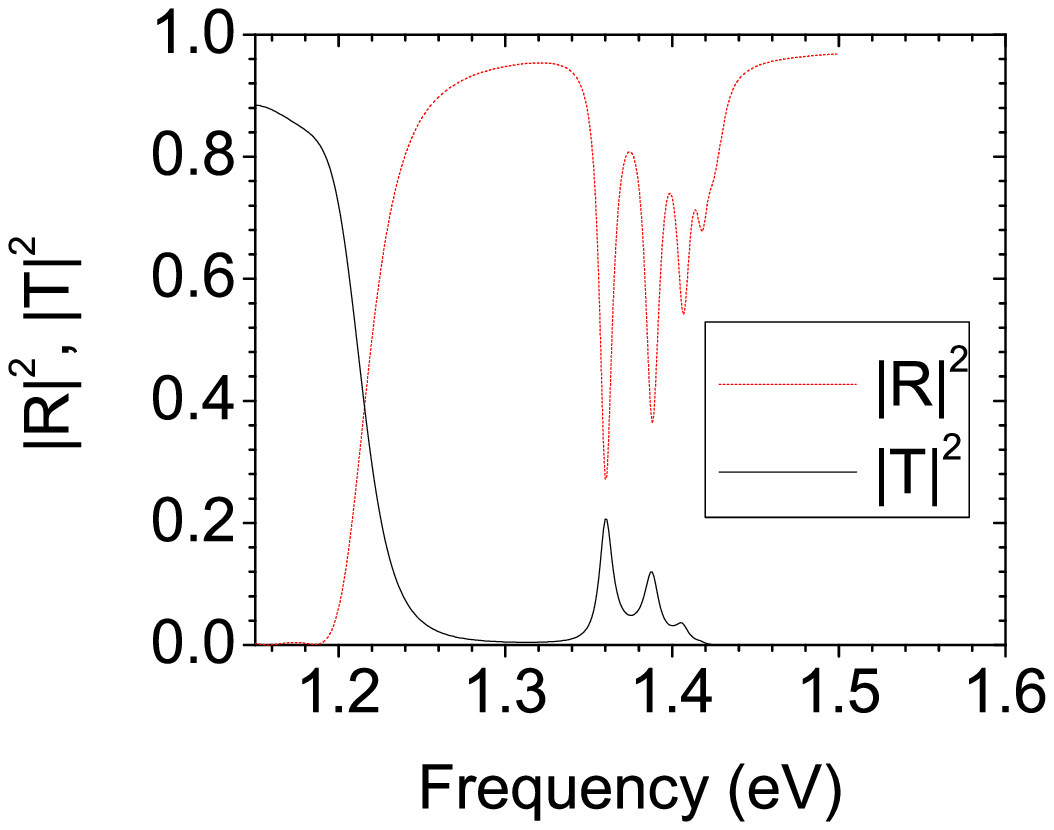}
\includegraphics[angle=-0,width=0.45\columnwidth]{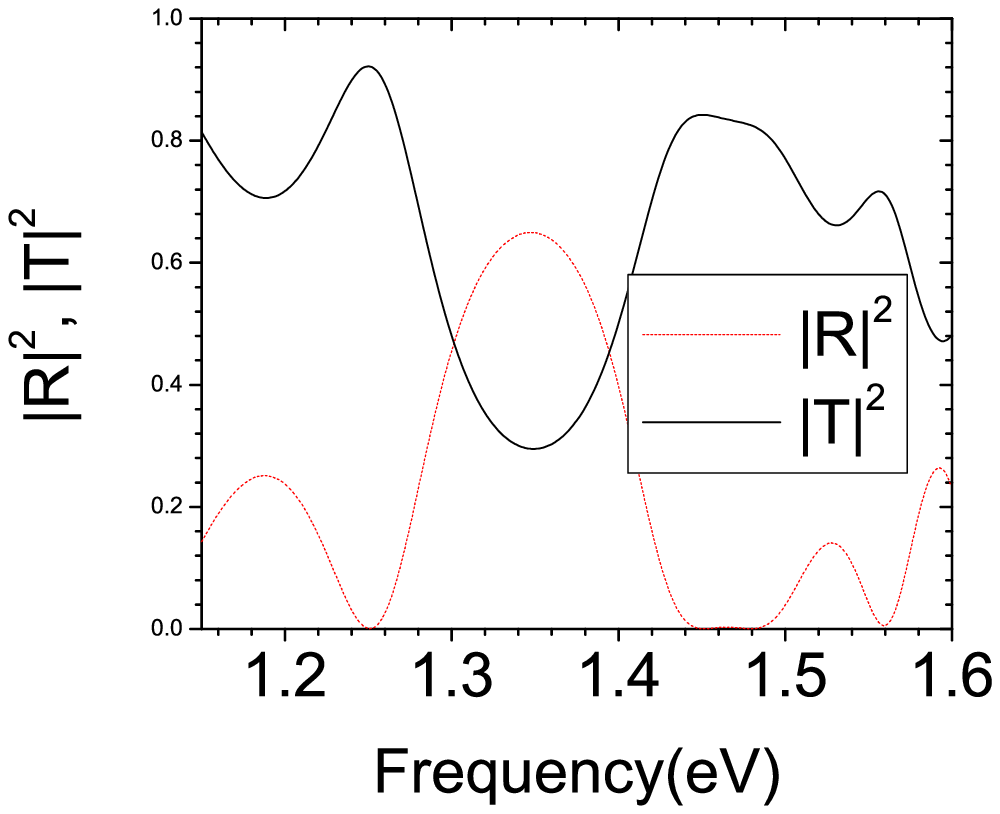}\\

\end{center}
\caption{(a), (b), (c), (d): Schematic diagram of the four SRR structures considered. The polarization and the direction of incidence of the radiation have been indicated.
(e), (f), (g), (h): Band structures for an array of SRRs of types (a), (b), (c) and (d) respectively. The solid circles indicate the real parts of the wave vector while the hollow circles denote the imaginary parts. The same colour (red or black) has been used for the real and imaginary parts of the eigenvalues with positive or negative real parts.
(i), (j), (k), (l): Reflection and transmission coefficients for a slab consisting of four layers of SRRs (a), (b), (c) and (d) respectively\\ }
\end{figure*}

It is clear from the figures that completely different behaviour is obtained in the three cases. The greatest surprise is the difference between structure(b) and structure(c) which is just the SRR in (b) reoriented. The most interesting new feature that comes up is that structure (a), structure(c) and structure(d) exhibit a negative refractive index in the second and third bands as can be seen from the fact that a waves with positive real wave number have a negative imaginary part of the wave number. The second band also has a negative group velocity which is not separated from the higher positive group velocity band by a band gap. In the case (a), these bands are both very flat and almost dispersionless. In case (b), we have in comparision only a bandgap with an avoided crossing with the next higher band. Interestingly, one obtains a gapless dispersion with a negative phase velocity  over a large band in case (d). These bands for (a), (c) and (d) have a negative phase velocity which can be seen from the fact that the real and imaginary parts of the wave-vector have opposite signs. Thus we have at literally optical frequencies (1.4 eV $\sim$ 882 nm) both the negative phase velocity  - negative group velocity as well as negative phase velocity and positive group velocity bands that were reported recently in an experiment on metamaterials\cite{may06_science}. Also note that at these frequencies, the free space wavelength of the light is about thrice the size of the unit cell and we have only one transmitted and reflected beam from a slab of such SRRs as all higher-order diffracted beams are evanescent.

The only identifiable difference between the systems is that the electric field of the incident radiation can interact with the charge distributions formed across two, nil or four capacitive gaps in the SRR (a), SRR (b) and SRR (c) respectively. This along with the fact that the usual SRR mechanism cannot be active at these high frequencies \cite{obrien_prb}, lead us to suspect that there are significant phase differences across a single SRR (b or c)  whose size $(\sim \lambda/3)$ becomes a significant fraction of the wavelength at the bandgaps.  The electric field would thus be able to interact with the dipoles which are not completely screened out due to retardation. To corroborate this, we computed the spectral problem for a small (3 X 4) array of SRR of type (c) using the COMSOL FEMLAB software with perfectly matched boundary layers (PML) placed at least more than 1000 nm away\footnote{We have a mismatch of about 0.02eV between the results obtained from FEMLAB and the transfer matrix method. The FEMLAB results give a frequency about 0.02 eV larger than the transfer matrix codes.  This could be due to differences in the discretization methods or the finite array size effects and PML conditons in the FEMLAB.}.For typical eigensolutions near 1.41eV, the magnetic fields were concentrated inside the SRRs as with the classical SRR, but pointed in opposite directions in adjacent layers. Similarly, the electric fields were concentrated in the capacitive gaps with large dipole like fields in the gaps which flip in sign across the SRR. The displacement fields go around the SRR similiar to a current. It could be concluded from the above that the electric field is exciting the anti-symmetric modes\cite{podolsky_jopa} in the vertical legs across the SRR with opposing currents (displacement and real) in the two legs. The interaction with the dipoles in the capacitive gaps is crucial to the easy excitation of the anti-symmetric mode. Thus the electric fields can effectively contribute to creating large magnetic fields under suitable conditions.

Further, in Fig. 2 and 3, we present our results on focussing a line source at 1.44eV frequency  by a finite flat slab made of SRR. This is the frequency where the light line intersects the negative group velocity band for the SRR (c).  Again the orientation of the SRR (b) or (c) turns out to be important, although in both cases we obtain an image. 
\begin{figure}[tb]
\begin{center}
\includegraphics[angle=-0,width=1.0\columnwidth] {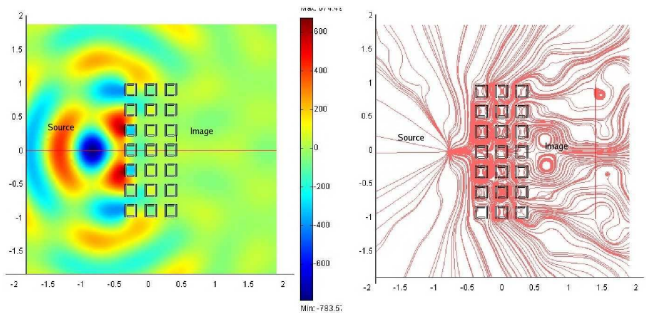}

\end{center}
\caption{ Left: Focussing a line source using a finite flat slab of SRR(b). The colour scheme indicates the intensity(arbitrary units) 
Right: Poynting vector map for the same slab}
\end{figure}
\begin{figure}[tb]
\begin{center}
\includegraphics[angle=-0,width=1.0\columnwidth] {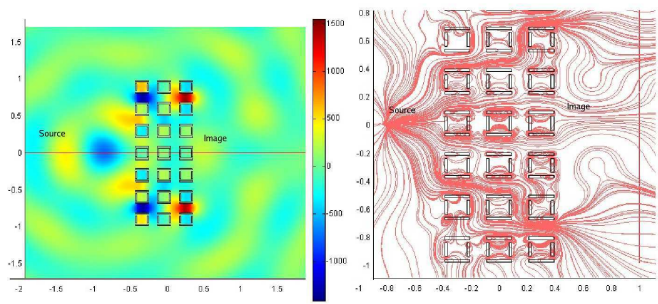}
\end{center}

\caption{Left: A line source focussed by a finite flat slab consisting of SRR(c). The colour scheme indicates the intensity(arbitrary units)
Right: Poynting vector map for the same slab}

\end{figure}

The usual tests of shifting the source around and putting in multiple sources were carried out to confirm the focussing effect. Subwavelength focussing effects by evanescent waves were not considered as the wavelength is quite comparable to the structure lengthscales. We have also verified that the imaging effects do not arise due to band-dispersion effects such as the all angle negative refraction (AANR)\cite{luo_aanr} by disordering the positions of the SRRs to upto 10\% of the lattice and also by removing one or two SRRs in the array. The images survive with marginal changes in the fields. 

In case of SRR(b), the slab transmits very little due to the bandgap, but there is a faint image on the other side (Fig. 2)which is  confirmed by the Poynting vector map which shows a focussing of the energy flow at the location of the image. In the case of SRR -(c) at the same frequency, the slab transmits a lot more and the image is also considerably brighter here (Fig. 3). This is also confirmed by a plot of  the Poynting vector map. But there are crucial differences in the two cases which enable us to make a distinction. For the SRR (b), the field at the image has the opposite sign of the field at the source point and the focus is at about 1.4 $\mu$m instead of about 0.9 $\mu$m as expected for the flat lens. This change of sign typically happens for imaging for p-polarized light with thick slabs of negative dielectric materials (See for example fig. 2(d) of Ref. \cite{shen_platzman} for the transmission coefficient), while the distance for the focus shown by the Poynting vector occurs somewhat further away than that predicted for the Veselago lens (See Fig. 3 of Ref. \cite{fang_apl2003}). Thus the array of SRR(b) is similiar to a stack of cut-wires (plates in this case), with the electric field of the incident radiation directed the plates(wires), which shows negative dielectric constant in the band-gap above the resonant frequency\cite{pendry96}. If this were the case, the capacitive gaps do not play much of a role in the phenomena. We verified this by closing the capacitive gaps and found that the weak imaging effect is essentially unchanged with hollow square cylinders as well. The band structure calculation for the hollow square cylinders shows a wide band-gap in the frequency range considered here, similar to SRR(b). Thus the whole system just behaves as a diluted metal and the imaging action for p-polarized light is due to a negative dielectric permittivity over a wide range of frequencies\cite{sar_jmo}. 

In case of SRR(c), we find that the images are brighter, the field at the image has the same sign as at the source position and also the image is found at almost exactly 0.9 $\mu$m where it is expected for the Veselago lens (Fig. 3(f)). Further it occurs where there is no bandgap. So we conclude that the focussing in this case (c) is due to negative refraction.
We have verified that the focussing happens in a reasonably wide range of frequencies 1.4 eV to 1.47eV while the best resolution is obtained at 1.44 eV. Due to the finite size and numbers of the SRRs used and dissipation we make no claims about the extent of the sub-wavelength resolution possible here. In the case of SRR(a) as well, we can see that there is a negative phase velocity band possible. 

We note an aspect of these systems that gives a clue to the negative refractive index at high frequencies. The large phase shifts for the radiation across a single unit cell implies that the electric dipole moments formed across the capacitive gaps will not be effectively screened out. In case (c), the electric field can also drive the currents around the loop by interacting with the dipole moments across the capacitive gaps and reinforces the effect due to the magnetic field. In  case (b), the electric dipole response of the finite wires (plates) dominates because the wires (plates) in the adjacent cells can couple through the electric field. This would explain the negative refractive index in case (c) and the negative dielectric medium in case(b). Thus, the system has plasmonic resonances of electric and magnetic character respectively, giving rise to the band gaps on either side of the negative phase velocity band. In case (d), the two plasmonic resonances appear to be degenerate, resulting in a gapless dispersion. We used the Green's tensor method to compute the magnetic dipole moment per SRR for a finite number of SRRs of finite length and found that a net magnetic dipole moment arises at resonance in both the case (b) and (c).

We do not believe that defining homogeneous effective medium parameters such as $\varepsilon$ and $\mu$ makes sense when homogenisation itself becomes questionable as the unit cell size is a significant fraction of the free space wavelength at resonance ($a \sim \lambda/3$). Hence we confine ourselves to the discussion of negative phase velocity or negative refractive index only for the shown incidence.

In conclusion, we have shown that symmetric SRRs can have negative phase velocity (negative refractive index) at optical frequencies if oriented such that the electric fields can drive the dipoles across the capacitive gaps. in the orthogonal orientation, they behave as cut-wire media with negative dielectric permittivity. The simple paradigm of a negative $\mu$ arising due to a L-C resonance in the SRR drive by the magnetic field alone breaks down and the negative refractive index that arise in the limits when homogenization becomes problematic have their origin in the plasmonic resonances of the system. Our theoretical work confirms the recent measurements of different combinations of positive and negative group velocity and negative phase velocity\cite{may06_science}. The results presented here give specific directions for the design of meta-materials with negative group and phase velocity (negative refractive index) at optical frequencies.

\section*{Acknowledgment}
SAR thanks S. Guenneau for helping with the calculations with FEMLAB and acknowledges partial support from the Department of Science and Technology, India under grant no.SR/S2/CMP-45/2003.

\end{document}